\newcommand{\size}[1]{\left\vert#1\right\vert}

\newcommand{\bigo}[1]{\mathcal{O}\left(#1\right)}
\newcommand{\bigT}[1]{\Theta\left(#1\right)}

\newcommand{\s}{\mathcal{F}}
\newcommand{\C}{\mathcal{C}}
\newcommand{\G}{\mathcal{G}}
\newcommand{\fie}{\mathbb{F}}
\newcommand{\weight}{\omega}

\newcommand{\Ba}[2][\delta]{\mathcal{B}_{#1}{\left({#2}\right)}}
\newcommand{\polylog}{\,\mathrm{polylog}}
\newcommand{\poly}{\,\mathrm{poly}}

\newcommand{\goal}{\emph{goal}}

\documentclass[letterpaper,11pt]{article}
\usepackage{amssymb,amsmath}
\usepackage[boxed]{algorithm2e}

\newtheorem{res}{Theorem}
\newtheorem{thm}{Theorem}
\newtheorem{lem}{Lemma}[section]
\newtheorem{defn}{Definition}[section]

\newcommand{\qedsymb}{\hfill{\rule{2mm}{2mm}}}
\newenvironment{proof}{\begin{trivlist}
\item[\hspace{\labelsep}{\sl\noindent Proof:\/}]
}{\qedsymb\end{trivlist}}

\newcommand{\Xomit}[1]{ }
\newcommand{\remove}[1]{}

\date{}

\numberwithin{equation}{thm}

\begin{document}
\title{Explicit Non-Adaptive Combinatorial Group Testing Schemes}

\author{Ely Porat\footnote{Bar-Ilan University, Dept. of Computer
Science, 52900 Ramat-Gan, Israel,
\emph{\texttt{porately@cs.biu.ac.il}}} \and Amir Rothschild $^*$
\footnote{Tel-Aviv University , Dept. of computer science,
Tel-Aviv, Israel, \emph{\texttt{rotshch@post.tau.ac.il}}}}


\maketitle 

\begin{abstract}
 Group testing is a long studied problem in combinatorics: A small
 set of $r$ ill people should be identified out of the whole ($n$ people) by using only
queries (tests) of the form ``Does set X contain an ill human?".
In this paper we provide an explicit construction of a testing
scheme which is better (smaller) than any known explicit
construction. This scheme has $\bigT{\min[r^2 \ln n,n]}$ tests
which is as many as the best non-explicit schemes have. In our
construction we use a fact that may have a value by its own right:
Linear error-correction codes with parameters $[m,k,\delta m]_q$
meeting the Gilbert-Varshamov bound
 may be constructed quite efficiently, in $\bigT{q^km}$ time.
\end{abstract}

\pagebreak
\section{Introduction}

Group testing is an important and well known tool in
combinatorics. Due to its basic nature, it has been found to be
applied in a vast variety of situations. In 2006 DIMACS has
dedicated a special workshop solely for the problem of group
testing \cite{dimacs}. 
A representative instance of group testing considers a set of
items, each of which can be either defective or non-defective, and
the task is to identify the defective items using the minimum
number of tests. Each test works on a group of items
simultaneously and returns whether or not that group contains at
least one defective item. A group testing algorithm is said to be
\emph{nonadaptive} if all the tests to be performed are specified
in advance. A formal definition is given in Section~\ref{PROBDEF}.

Group testing has a long history dating back to at least
1943~\cite{Dorfman:1943}. In this early work the problem of
detecting syphilitic men for induction into the United States
military using the minimum number of laboratory tests was
considered. While this idea is still relevant today for testing
viruses such as HIV, it is only one of the many applications found
for group testing: In the effort of mapping genomes, for example,
we have a huge library of DNA sequences, and test whether each of
them contains a probe from a given set of DNA pieces
\cite{BLC,BBKB,Berger:S}. Somewhat less conventional uses for
group testing were introduced lately in pattern matching
algorithms
\cite{DBLP:conf/esa/CliffordEPR07,DBLP:conf/cpm/AmirKP07} and in
streaming algorithms \cite{1061325}: For instance,
\cite{DBLP:conf/esa/CliffordEPR07} solves the problem of searching
for a pattern in a text with a bounded number of mismatches. A
recent paper about pattern matching in a streaming model even
utilizes group testing twice in the same algorithm \cite{PR09}.
Additional applications of group testing include: compressed
sensing
\cite{muth0685,ind0885,1250824,DBLP:conf/sirocco/CormodeM06,DBLP:journals/corr/abs-0708-1211},
quality control in product testing \cite{SG}, searching files in
storage systems \cite{KS}, sequential screening of experimental
variables \cite{LI}, efficient contention resolution algorithms
for multiple-access communication \cite{KS,Wolf85}, data
compression \cite{ hong00group}, software testing \cite{BlassG02,
cohen97aetg}, DNA sequencing \cite{PevznerL94} and other
applications in computational molecular biology
\cite{DH:2000,FKKM,ND,1996IMA....81..133B}. In most of the
algorithms and applications presented here, our group testing
algorithm generates improvements to the results.

Consider the situation where there are $n$ items out of which at
most $r$ are defective. It has been shown that in this situation
any nonadaptive combinatorial group testing ($(n,r)$-GT) procedure
must use $\Omega(\min[r^2\log_r{n},n])$ tests
\cite{chaudhuri96deterministic}. The best known schemes use
$\bigT{\min[r^2\ln{n},n]}$ tests \cite{KS}, and the best known
explicit (polynomial time constructable) schemes need as much as
$\bigT{\min[r^2\log_{r\ln n}^2{n},n]}$ tests \cite{KS}. In this
paper, we present an explicit GT scheme which contains merely
$t=\bigT{\min[r^2\ln{n},n]}$ tests (the same as the best known
non-explicit schemes), and takes $\bigT{rn\ln{n}}$ time to build,
which is linear in its representation ($\bigo{t\frac{n}{r}}$).
Hence, this paper closes the gap between the explicit and
non-explicit group testing schemes.

\subsection{Error Correction Codes}

An error-correcting code (ECC) is a method for encoding data in a
redundant way, such that any errors which are introduced can be
detected and corrected (within certain limitations). Suppose Alice
wants to send Bob a string of $k$ letters from an alphabet of size
$q$ using some noisy channel. An $(m,k,d)_q$  error-correction
code enables Alice to encode her string to an $m>k$ letters
string, such that Bob will be able to detect whether the received
message has up to $d$ errors, and even decode the message if it
has less than $\frac{d}{2}$ errors. A linear code (LC) is an
important type of error-correction code which allows more
efficient encoding and decoding algorithms than other codes.
Error-correction codes are used in a vast variety of fields
including information transmission, data preservation,
data-structures, algorithms, complexity theory, and more.

One of the most important goals of coding theory is finding codes
that can detect many errors, while having little redundancy. The
Gilbert-Varshamov (GV) bound shows this can be done to some
extent: We define the \emph{rate} of a code, $R = \frac{k}{m}$ and
the \emph{relative distance} of a code, $\delta = \frac{d}{m}$.
The GV bound asserts that there are codes with $R\geq
1-H_q(\delta)-o(1)$ where $H_q(p)$ is the $q$-ary entropy function
$$H_q(p)=p\log_q{\frac{q-1}{p}}+(1-p)\log_q{\frac{1}{1-p}}$$ and
$o(1)\xrightarrow[m\rightarrow\infty]{} 0$ \cite{GIL:52,VAR}.
Though the GV bound is half a century old, no explicit
construction of codes meeting it has yet been found. The best
known construction takes polynomial time in $q^{m-k}$
\cite{BrualdiP93}.

We present a more efficient deterministic construction for linear
codes meeting the GV bound. Our construction takes $\bigT{q^km}$
time. The importance of this result is apparent when constructing
codes with low rates; First, for small rates the GV bound is the
best known lower bound on the rate and relative distance of a
code. Second, the lower the rate, the slower the previously known
best construction, and the faster our construction.

\subsection{Previous Results}

Since the problem of group testing was first introduced in 1943,
many problems related to it and generalizations of it were
considered including: fully-adaptive group testing , two staged
group testing and selectors \cite{chrobak00fast,knill95lower,
BGV:2003,
DBLP:conf/soda/ClementiMS01,DBLP:journals/tcs/BonisV03,DBLP:journals/siamcomp/BonisGV05},
group testing with inhibitors
\cite{830034,DBLP:journals/ipl/Damaschke98,DBLP:journals/ipl/BonisV98,BGV:2003},
group testing in a random case where a distribution is given on
the searched set \cite{knill95lower,
journals/tit/BergerL02,861005, Macula:1998:1382-6905:385,
BGV:2003}, group testing in the presence of errors
\cite{DBLP:journals/dam/KnillBT98} and more. Regarding the
original problem of group testing, Kautz and Singleton \cite{KS}
proved the existence of GT schemes of size $\bigT{r^2\ln{n}}$, and
showed how to explicitly construct schemes of size
$\bigT{\min[r^2\log_{r\ln n}^2{n},n]}$. They also managed to give
an explicit construction of schemes of size $\bigT{\ln n}$ for the
special case $r=2$. Since their work, no asymptotic improvements
to the size of the GT scheme were found. One paper succeeded,
however, in improving the size of the \emph{explicit} schemes (but
only for \emph{constant} values of $r$):
\cite{DBLP:journals/talg/AlonMS06} showed how to construct an
explicit construction of schemes of size $\bigT{r^2\ln n}$ in time
polynomial in $n^r$. From the probabilistic perspective, there is
no known Las-Vegas algorithm (though one easily stems from our
methods) constructing a scheme of size $\bigT{r^2\ln n}$. The only
known probabilistic constructions are Monte-Carlo algorithms.

Regarding error-correction codes the picture is more complex. The
GV bound was first presented by Gilbert in 1952 \cite{GIL:52}. He
provided a $\bigT{q^m}$ time greedy construction for codes meeting
his bound. A few years later Varshamov \cite{VAR} showed linear
codes share this bound and Wozencraft \cite{WOR:63} offered a
$\bigT{q^m}$ time deterministic construction of such codes. In
1977 Goppa \cite{Goppa:77} initiated the fruitful study of
algebraic geometric codes. Codes eventually found by this study
surpass the GV bound for various alphabet sizes and rates
\cite{TVZ:82}. Recently, an explicit, $\bigT{m^3\polylog{m}}$ time
construction was given for algebraic geometric codes
\cite{Shum:2000,Shum:2001}. The best deterministic construction
for alphabet sizes and rates where the GV bound is superior to the
algebraic geometric bound, was provided in 1993 by Brualdi and
Pless \cite{BrualdiP93}. They presented a $\poly(q^{m-k})$
construction of binary linear codes meeting the GV bound. Their
construction can be easily generalized to deal with larger
alphabets. Even under hardness assumptions, no explicit
construction of codes meeting the GV bound has yet been found,
though an effort presenting some worthy results is given in
\cite{ALGO-CONF-2007-002}.

\subsection{Our Results}
We present the first explicit $(n,r)$-GT scheme which contains $t
= \bigT{\min[r^2\ln{n},n]}$ tests, thus closing the gap between
explicit and non-explicit group testing schemes. Our construction
takes $\bigT{rn\ln{n}}$ time to build, meaning linear time in its
representation ($\bigo{t\frac{n}{r}}$).
\begin{res}
Let $n$ and $r$ be positive integers. It is possible to construct
a $(n,r)$-GT containing $\bigT{\min[r^2\ln n,n]}$ tests in
$\bigT{rn\ln n}$ time.
\end{res}

We also present the most efficient deterministic construction for
linear codes meeting the GV bound. Our construction builds an
$[m,k,\delta m]_q$-LC in $\bigT{q^km}$ time.
\begin{res}
Let $q$ be a prime power, $m$ and $k$ positive integers and
$\delta \in [0,1]$. If $k \leq \left(1-H_q(\delta)\right)m$, then
it's possible to construct an $[m,k,\delta m]_q$-LC in time
$\bigT{mq^k}$.
\end{res}

\subsection{The Paper Outline}
We start this paper with formal definitions in
Section~\ref{PROBDEF}, and continue by showing a connection
between error-correction codes and group testing schemes in
Section~\ref{ECCSSF}. Then we immediately move to the main result
of the paper in Section~\ref{EXPSSF}, showing how to efficiently
construct small group testing schemes. This construction for group
testing schemes uses our construction of a linear code which is
given in Section~\ref{GVB}.

\section{Problems Definitions}\label{PROBDEF}
\begin{defn}
Consider a universe $U$. A family of tests (subsets) $\s \subset
\mathcal{P}(U)$ is a group testing scheme of strength $r$
($(n,r)$-GT) if for any subset $A \subset U$ of size at most $r$,
and for any element $x \notin A$, there exist a test $B \in \s$
that distinguishes $x$ from $A$, meaning $x\in B$ while $A \cap B
= \varnothing$.
\end{defn}

In order to ease reading, we present short notations of an
error-correction code and a linear code.
\begin{defn}
An ECC, $\C$, is said to have parameters $(m,k,d)_q$ if it
consists of $q^k$ words of length $m$ over alphabet $\Sigma$ of
$q$ elements, and has Hamming distance $d$. Such an ECC is denoted
as $(m,k,d)_q$-ECC.
\end{defn}

\begin{defn}
An $[m,k,d]_q$-LC is a special case of an $(m,k,d)_q$-ECC which is
over alphabet $\Sigma=\fie_q$ when the codewords form a linear
subspace over $\fie_q^m$. Such a linear code is said to have
parameters $[m,k,d]_q$. A linear code has a \emph{generator
matrix} $\G \in \mathcal{M}_{m\times k}$ which generates it,
meaning $\C = \{\G y \mid y \in \fie_q^k \}$.
\end{defn}

\section{Background}\label{ECCSSF}
Our results concerning GT are more natural and straightforward
using the combinatorial concepts \emph{selection by intersection}
and \emph{strongly-selective family} (SSF) \cite{CMS}. Selection
by intersection means distinguishing an element from a set of
elements by intersecting it with another set. More precisely,
\begin{defn}
Given a subset $A \subset U$ of a universe $U$, element $x \in A$
is \emph{selected} by subset $B \subset U$ if $A \cap B = \{x\}$.
An element is \emph{selected} by a family of subsets $\s \subset
\mathcal{P}(U)$ if one of the subsets in $\s$ selects it.
\end{defn}
An SSF is a family of subsets that selects any element out of a
small enough subset of the universe. More precisely,
\begin{defn}
A family $\s \subset \mathcal{P}(U)$ is said to be
\emph{$(n,r)$-strongly-selective} if, for every subset $A \subset
U$ of size $\size{A}=r$, all elements of $A$ are selected by $\s$.
We call such a family an $(n,r)$-SSF.
\end{defn}

SSFs and GT schemes are strongly connected: On the one hand, an
$(n,r+1)$-SSF is a GT scheme of strength $r$, and on the other
hand, a GT scheme of strength $r$ in a universe of size $n$ is an
$(n,r)$-SSF. For a detailed proof see \cite{KS}.


In what follows we will focus on SSF constructions. It is
important to note that explicit constructions for SSFs give
explicit constructions for GT schemes with the same asymptotic
behavior. Next we show how to construct an SSF from an ECC, and
how good this construction is. The foundations of the idea we
present was developed in an earlier work by Kautz and Singleton on
superimposed codes \cite{KS}. The context and formalisms that were
employed are quite distinct from those we require, the idea is
quite simple and though, we are not aware of this aspect of their
work being developed subsequently. Thus, the following subsection
will provide full and complete explanations and proofs of the
construction.

\subsection{Reducing ECCs to SSFs}\label{ECCSSF:reduction}

As it turns out, one can build small strongly-selective families
from good error-correction codes having large distance. Both the
construction and the proof are given in this Subsection. In a few
words, the idea behind the construction is that taking a small set
of codewords from the ECC and another codeword $w$, there must be
positions in which $w$ differs from all the words in this set.
This is because $w$ differs from any other word in the code in
many positions, and so, in a small set of codewords, there must be
some shared positions in which all codewords differ from $w$.
Therefore we'll get an SSF if we first translate elements of $[n]$
to codewords, and second, find tests which isolate a codeword $w$
from a set of codewords if it differs from this set in a certain
position. We construct such tests by assembling a test for each
possible letter in each possible position in the word. A detailed
construction follows.

 Suppose $\C =
\{w_1,...,w_n\}$ is an $(m,\log_q{n},\delta m)_q$-ECC. The
constructed SSF, $\s(\C)$, will be assembled from all the sets of
indexes of codewords that have a certain letter in a certain
position. More accurately, for any $p \in [m]$ and $v \in [q]$,
define $s_{p,v}=\left\{i \in [n] \mid w_i[p]=v \right\}$. Define
$\s(\C)$ as the set of all such $s_{p,v}$-s: $\s(\C)=\left\{
s_{p,v} \mid p \in [m] \;and\; v \in [q] \right\} $.

The size of $\s(\C)$ is at most $mq$. Notice that this
construction may be performed in time $\bigT{nm}$ (linear in the
size of the representation of $\s$) using
Algorithm~\ref{ECCSSF:reduction:alg}.

\begin{algorithm}[H]\caption{Constructing an SSF from an ECC}\label{ECCSSF:reduction:alg}
\SetLine \SetVline
 \ForEach{$i \in [n]$} {
  \ForEach{$p \in [m]$} {
   insert $i$ into $s_{p,w_i[p]}$ \;
  }
 }
\end{algorithm}

The following Lemma shows that this construction really does
result in a small SSF, and more specifically, that $\s(\C)$ is an
$(n,\lceil\frac{1}{1-\delta}\rceil)$-SSF.
\begin{lem} \label{ECCSSF:lemGG}
Let $\C$ be an $(m,\log_qn, \delta m)_q$-ECC. Then $\s(\C)$ is an
$(n,\lceil\frac{1}{1-\delta}\rceil)$-SSF.
\end{lem}
\begin{proof}
Let $r=\lceil\frac{1}{1-\delta}\rceil$. Let $i_1,...,i_r \in [n]$
be any $r$ distinct indexes in $[n]$. W.L.O.G. we prove that $i_1$
is selected from $\{i_1,...,i_r\}$ by $\s(\C)$. For any $j \neq
1$, the number of positions $p\in [m]$ where
$w_{i_j}[p]=w_{i_1}[p]$ is at most $(1-\delta) m$. Thus, the
number of positions where
$w_{i_1}[p]\in\{w_{i_2}[p],...,w_{i_r}[p]\}$ is at most $(r-1)
(1-\delta) m < m$. Therefore, there exist a position $p$ where
$w_{i_1}[p]\notin\{w_{i_2}[p],...,w_{i_r}[p]\}$. This means that
$i_1\in s_{p,w_{i_1}[p]}$ while all other $i_j$-s are not. Thus,
$i_1$ is selected by $s_{p,w_{i_1}[p]}$. 
\end{proof}

For illustration, we consider the following example: If we test
our algorithm on the Reed-Solomon $[3,2,2]_3$-LC:
$$\C=\{\texttt{000, 111, 222, 012, 120, 201, 021, 102, 210} \}$$
We get the following $(9,3)$-SSF:
\begin{align*}
\s(\C)=\{\quad&\texttt{\{1,4,7\}, \{2,5,8\}, \{3,6,9\},} \\
&\texttt{\{1,6,8\}, \{2,4,9\}, \{3,5,7\},} \\
&\texttt{\{1,5,9\}, \{2,6,7\}, \{3,4,8\}} \quad\}
\end{align*}
\section{Main Theorem} \label{EXPSSF}

\begin{thm}\label{EXPSSF:corSMALL}
Let $n$ and $r$ be positive integers. It is possible to construct
an $(n,r)$-SSF of size $\bigT{\min[r^2\ln n,n]}$ in $\bigT{rn\ln
n}$ time.
\end{thm}

\begin{proof}
If $r^2\ln n \geq n$, simply return the $n$ tests $\{i\}_{i=1}^n$.
We continue the proof assuming that $r^2\ln n < n$. Set
$\delta=\frac{r-1}{r}$ (which is equivalent to
$r=\frac{1}{1-\delta}$), $q \in [2r,4r)$ a prime power, $k=\log_q
n$ and $m=\frac{k}{1-H_q(\delta)}=\bigT{kr\ln r}=\bigT{r\ln n}$.

Use Theorem~\ref{GVB:DET:thmFinal} to construct an $[m,k,\delta
m]_q$-LC in time $\bigT{nm}$. This is possible since $k \leq
\left(1-H_q(\delta)\right)m$.

According to Lemma~\ref{ECCSSF:lemGG}, we can now construct an
$(n,r)$-SSF of size $mq=\bigT{r^2\ln n}$. The time this
construction will take is $\bigT{nm}=\bigT{rn\ln n}$.
\end{proof}

\section{Meeting the Gilbert-Varshamov bound more Efficiently}\label{GVB}

In this Section we demonstrate a deterministic construction of LCs
which meets the GV bound. We developed this deterministic
algorithm by taking a randomized algorithm and derandomizing it
using the method of conditional probabilities (a full discussion
concerning this method is given in \cite{ProbMethod}). Using this
method requires the randomized algorithm to have several
non-trivial attributes. First, there need to be a goal function
$\emph{goal}: \emph{LinearCodes}\rightarrow \mathbb{R}$ which
returns a large result whenever the randomized algorithm fails.
Second, this function has to have low expectation - lower than the
minimum value returned by it when the algorithm fails. Third, the
random selections of the algorithm have to be divided into stages
with a small number of options to choose from in each. Finally,
there should be an efficient algorithm for calculating in each
stage of the algorithm the option minimizing the expectation of
$\goal$ given all the selections done until that point. In
Subsection~\ref{GVB:prob} we'll show the randomized algorithm,
present the goal function $\goal$, show that the algorithm fails
iff $\goal(\G)\geq 1$ (where $\G$ is the generator matrix returned
by the algorithm), and show that $E(\goal)<1$. In
Subsection~\ref{GVB:DET} we'll present the derandomized algorithm
more accurately, showing how to divide it to the small stages.
We'll also prove it should work, and show how to calculate the
option minimizing the expectation of $\goal$ in each stage. We'll
finish this Subsection having an algorithm taking time polynomial
in the complexity we desire, we improve it in
Subsection~\ref{GVB:IMP} to acquire the desired complexity.

\subsection{The Probabilistic Algorithm}\label{GVB:prob}
Algorithm~\ref{GVB:prob:AlgoProb} is a standard probabilistic
algorithm for building linear codes with rate meeting the GV
bound.

\begin{algorithm}[H]\caption{Probabilistic Construction of a Linear Code}\label{GVB:prob:AlgoProb}\SetLine \SetVline
\KwIn{$m,k \in \mathbb{N}$, $\delta \in [0,1]$ s.t. $k \leq
(1-H_q(\delta))m$} Pick entries of the $m\times k$ generator
matrix $\G$ uniformly and independently at random from $\fie_q$\;
\KwOut{$\G$}
\end{algorithm}

\begin{defn}
Given a codeword $x$ of length $m$, and a distance parameter
$\delta \in [0,1]$, we define $\Ba{x}$ as the bad event that the
weight of $x$ is less than $\delta m$, $\weight (x)<\delta m$. By
abuse of notation we refer to $\Ba{x}$ also as the indicator of
the same event.
\end{defn}
If we manage to choose a code with no bad event (not considering
the 0 codeword, of course), then the weight of the generated code
is larger than $\delta m$. As the weight and distance of a linear
code are equal, the algorithm succeeds. Therefore, our goal
function will be $\goal(\G) = \sum_{0 \neq y \in \fie_q^k}\Ba{\G
y}$. The algorithm succeeds iff $\goal(\G)=0$. We now need to show
that $E(\goal)$ is small. Therefore, we are interested in proving
that the probability of a bad event is sufficiently small. In
order to do so, we use the following version of the Chernoff
bound:

\begin{thm}[Chernoff bound \cite{chernoff}]

Assume random variables $X_1,...,X_m$ are i.i.d. and $X_i \in
[0,1]$. Let $\mu=E(X_i)$, and $\epsilon > 0$. Then

$$Pr \left( \frac{1}{m}\sum{X_i} \geq \mu+\epsilon \right) \leq \left( \left( \frac{\mu}{\mu+\epsilon} \right)^{\mu+\epsilon} \left( \frac{1-\mu}{1-\mu-\epsilon} \right)^{1-\mu-\epsilon} \right)^m=e^{-D(\mu+\epsilon||\mu)m}$$ where $D(x||y)=x\log\frac{x}{y} + (1-x)\log\frac{1-x}{1-y}$.
\label{thm:chernoff1}
\end{thm}

\begin{lem}\label{GVB:prob:lemBadSmallProb}
Let $y$ be a nonzero vector in $\fie_q^k$. Let $\G$ be a random
generator matrix chosen according to
algorithm~\ref{GVB:prob:AlgoProb}. Then
$\log_q\left(Pr\left(\Ba{\G y}\right)\right) \leq
-m\left(1-H_q\left(\delta\right)\right)$.
\end{lem}

\begin{proof}
It is easy to see that $x=\G y$ is a random vector in $\fie_q^m$.
Therefore, $\weight (x)$ is binomially distributed; $\weight (x)
\sim B\left(m,1-\frac{1}{q}\right)$. 
Using the Chernoff bound (Theorem~\ref{thm:chernoff1}) we get
$$Pr\left(\Ba{x}\right)=Pr\left(\weight(x) \leq \delta m\right)\leq \left(   \left(\frac{\frac{1}{q}}{1-\delta} \right)^{1-\delta}   \left( \frac{1-\frac{1}{q}}{\delta}  \right)^{\delta}  \right)^m$$
Extracting logarithm from the former expression and simplifying it
we attain
\begin{eqnarray*}
  \log_q\left(Pr\left(\Ba{x}\right)\right) &\leq& m \left( \left(1-\delta\right)\left(-1-\log_q\left(1-\delta\right)\right)+\delta\left(\log_q\left(1-\frac{1}{q}\right)-\log_q{\delta}\right)   \right) \\
  &= & -m\left(1-H_q\left(\delta\right)\right)
\end{eqnarray*}
\end{proof}

We will now show that for an appropriate choice of parameters, the
expected number of bad events, $E(\goal)$, is smaller than 1.
\begin{lem} \label{GVB:prob:lemFewBad}
Suppose $\G$ is a random generator matrix chosen according to
algorithm~\ref{GVB:prob:AlgoProb}. Suppose that $k \leq
(1-H_q(\delta))m$. Then $E(\goal) < 1$.
\end{lem}

\begin{proof}

By linearity of the expectation
$$E(\goal)=E\left(\sum_{y \neq 0}\Ba{\G y}\right) =\sum_{y \neq 0}E\left(\Ba{\G y}\right) =\sum_{y \neq 0}Pr\left(\Ba{\G y}\right)$$
Next, employ Lemma~\ref{GVB:prob:lemBadSmallProb} to acquire that
$$E(\goal) \leq (q^k-1)q^{-m\left(1-H_q\left(\delta\right)\right)} < q^{k-m\left(1-H_q\left(\delta\right)\right)}$$
And finally, use our assumption $k \leq (1-H_q(\delta))m$ to
achieve the desired result
$$E(\goal)  < 1$$
\end{proof}

\subsection{Derandomizing the Algorithm}\label{GVB:DET}

Next we will show how to derandomize the algorithm.
Algorithm~\ref{GVB:DET:AlgFind} will determine the entries of the
generator matrix one by one, while trying to minimize the
expectation of the number of bad events, $\goal$.

\begin{algorithm}[H]\caption{Finding a code having no Bad
Events}\label{GVB:DET:AlgFind} \SetLine \SetVline \KwIn{$m,k \in
\mathbb{N}$, $\delta \in [0,1]$ s.t. $k \leq (1-H_q(\delta))m$}
Initialize $\G$ to be an $m \times k$ matrix\; \ForEach{$i \in
[m]$} {
 \ForEach{$j \in [k]$} {
  Set $\G[i,j]$ so as to minimize the expected value of $\goal(\G)$
 given all the values of $\G$ chosen so far\;
 }
} \KwOut{$\G$}
\end{algorithm}

Two questions arise from the above description of the algorithm:
First, will this algorithm find a code with no bad events? Second,
how can we find the value of $\G[i,j]$ in each step of the
algorithm?

The answer to the first question is, of course, positive. The
presented algorithm works according to the derandomization scheme
of conditional probabilities, and so, the number of bad events in
the returned solution will be no more than the expectation of this
number before fixing any of the letters. We'll delve into the
proof after introducing some additional notations concerning the
algorithm:

\begin{defn}
We assert that the algorithm is in \emph{step-$(i,j)$} when it is
about to choose the entry $(i,j)$ in $\G$. We denote the step
following $(i,j)$ by $(i,j)+1$.
\end{defn}
\begin{defn}
$ST_{(i,j)}$ will denote the state of the matrix $\G$ at step
$(i,j)$ -- i.e. which entries have been fixed to which values.
\end{defn}

\begin{lem}\label{GVB:DET:lemFewBad}
The above algorithm will find a code with no bad events, i.e.
$\goal(\G)=0$.
\end{lem}
\begin{proof}
Suppose the algorithm is in some step $(i,j)$.
$$Pr(\Ba{\G y} \mid ST_{(i,j)}) = \frac{1}{q}\sum_{v \in \Sigma}Pr(\Ba{\G y}
\mid ST_{(i,j)}\;,\;\G[i,j]=v)$$ Consequently,
\begin{eqnarray*}
  E(\goal \mid ST_{(i,j)}) &=&E(\sum_{y\neq 0}\Ba{\G y}\mid ST_{(i,j)}) \\
  &=&\sum_{y\neq 0}Pr(\Ba{\G y} \mid ST_{(i,j)}) \\
  &=&\frac{1}{q}\sum_{v \in \Sigma}\sum_{y\neq 0}Pr(\Ba{\G y} \mid
ST_{(i,j)}\;,\;\G[i,j]=v) \\
  &\geq&\min_{v\in\Sigma}\sum_{y\neq 0}Pr(\Ba{\G y}  \mid
ST_{(i,j)}\;,\;\G[i,j]=v) \\
  &=&\sum_{y\neq 0}Pr(\Ba{\G y}  \mid  ST_{(i,j)+1}) \\
  &=&E(\sum_{y\neq 0}\Ba{\G y} \mid  ST_{(i,j)+1}) \\
  &=&E(\goal \mid ST_{(i,j)+1})
\end{eqnarray*}

Therefore, if the values of the entries are chosen one by one, so
as to minimize the expectation of $\goal$, this value can not
increase. Since this value is smaller than $1$ in the beginning
according to Lemma~\ref{GVB:prob:lemFewBad}, it follows that it is
smaller than $1$ in the end. But at the end all entries are
chosen, and hence the value of $\goal$ will be exactly the number
of bad events that hold for the codewords we have chosen. This
number must be an integer, hence, it is 0.
\end{proof}
The answer to the second question, regarding how to find what the
value of $\G[i,j]$ should be, requires additional work. It would
be convenient to order the vectors $y \in \fie_q^k$ according to
the lexicographic order, setting $y_\ell$ to be the $\ell$-th
vector according to the lexicographic order.

We need to know for any codeword the number of positions in which
it vanishes, at each step of the algorithm. For this purpose
maintain an array $A$ of $q^k$ entries throughout the algorithm.
Entry $A[\ell]$ in this array will hold the number of positions in
which the code-word $\G y_\ell$ vanished so far. Maintaining this
array will require overall $\bigT{mq^k}$ time. This is due to the
fact that in each step $(i,j)$ we only need to consider changing
the values $A[y_\ell]$ for $q^{j-1}\leq \ell < q^j$ since the only
letters we fixed during this step belong to these words. We claim
that the number of position where the word $\G y_\ell$ vanishes
determines the conditioned probability of $\Ba{y_\ell}$.

\begin{lem}\label{GVB:DET:BadBinom}
Consider a codeword $\G y_\ell$ for which  all entries up to $i$
were fixed (by the entries selected in $\G$), and entries $i$ to
$m$ were not fixed yet. In other words, there exists a word $f \in
\fie_q^{i}$ of length $i$, such that for each possible $\G$, and
$\forall t \leq i:(\G y_\ell)[t] = f[t]$, and the same is not true
for $i+1$. Also suppose that until now, $\G y_\ell$ doesn't vanish
on exactly $c$ positions ($c=\size{\{t \leq i \mid f[t]\neq
0\}}$). Then $\weight(\G y_\ell)-c \sim B(m-i , 1-\frac{1}{q})$,
and $Pr(\Ba{\G y_\ell} \mid \forall t \leq i:(\G y_\ell)[t] =
f[t])$ is the probability that such a binomial variable will be
smaller than $\delta m - c$.
\end{lem}

\begin{proof}
Any entry which wasn't fixed in $\G y_\ell$, has a probability of
$1-\frac{1}{q}$ to vanish. The entries in $\G y_\ell$ are
independent of one another, and thus, $\weight(\G y_\ell)-c \sim
B(m-i , 1-\frac{1}{q})$.
\end{proof}

Now, in step $(i,j)$, For any codeword $\G y_\ell$ s.t. $q^{j-1}
\leq \ell < q^j$, we can calculate the probabilities
$Pr(\Ba{y_\ell} \mid ST_{(i,j)}\;,\;\G[i,j]=v)$ for all $v\in [q]$
in $\poly(q^k,m)$ time using Lemma~\ref{GVB:DET:BadBinom}.
Consequently, we can calculate all the expectations
$E(\sum_{q^{i-1}\leq \ell < q^i}\Ba{y_\ell} \mid
ST_{(i,j)}\;,\;G[i,j]=v)$ for all $v\in [q]$ in $\poly(q^k,m)$
time and find the value of $v$ which minimizes this expectation.
Hence, we can complete Algorithm~\ref{GVB:DET:AlgFind} in
$\poly(q^k,m)$ time. In the following Subsection we give
improvements to this algorithm, showing how to achieve the desired
complexity.

\subsection{Improving the Deterministic Algorithm} \label{GVB:IMP}

In order to find the letter $v$ which minimizes $E_{i,j,v} =
E(\sum_{q^{j-1}\leq \ell < q^j}\Ba{y_\ell} \mid
ST_{(i,j)}\;,\;\G[i,j]=v)$, we do not actually have to calculate
the $q$ expectations $E_{i,j,v}$. It is enough to calculate the
differences of those expectations and a constant value. We will
use the constant value which is the expected number of bad events
given $ST_{(i,j)}$ and that $(\G y_\ell)[i] \neq 0$ for all
$q^{j-1}\leq \ell < q^j$ (Of course, it's improbable that no
letters would vanish in step $(i,j)$, as the purpose of this
assumption is only to help us with the proof). We denote this
constant value $E_{i,j}$.

According to Lemma~\ref{GVB:DET:BadBinom}, for any vector $y$ the
following holds:
\begin{eqnarray*}
  &Pr(\Ba{\G y} \mid ST_{i,j}\;,\;(\G y)[i]=0) - Pr(\Ba{\G y}  \mid  ST_{i,j}
\; , \; (\G y)[i] \neq 0) =  \\
&{{m-i} \choose {\delta m-c}}\left(1-\frac{1}{q}\right)^{\delta m
-c} \left(\frac{1}{q}\right)^{(m-i)-(\delta m -c)}
\end{eqnarray*}
Denote the above expression $\mathrm{Dif}_{i,j}(y)$. Let $T$ be
the time it takes to calculate this expression. Now, we can
calculate all $q$ differences $E_{i,j,v}-E_{i,j}$ quite
efficiently in the following manner: Initialize a size $q$ array
$W$. Then, run over the vectors $y_\ell$ for $q^{j-1} \leq \ell <
q^j$, and for each subtract the difference
$\mathrm{Dif}_{i,j}(y_\ell)$ from cell
$v=-y_\ell[j]^{-1}\sum_{t=0}^{j-1}\G[i,t]y_\ell[t]$ in $W$ (since
this cell means setting $(\G
y_\ell)[i]=\sum_{t=0}^{j-1}\G[i,t]y_\ell[t]+\G[i,j]y_\ell[j]=0$).
After considering all values of $y_\ell$ , the position with the
maximal value in $W$ is the letter we should set for $\G[i,j]$.
Each entry number $v$ can be calculated in constant time for all
$q^{j-1} \leq \ell < q^j$ if we traverse over the $\ell$-s in each
step according to Gray code. Overall, the program will calculate
$mq^k$ entries, and so, it will take $mq^kT$ time.

Finally, we will show how to drop the $T$ factor and achieve a
$\bigT{mq^k}$ running time. In order to do so we need to take two
measures:
\begin{itemize}
    \item Use standard approximation techniques throughout the algorithm to
approximate the weights in $W$ instead of calculating them
exactly. 
    \item Evaluate approximately all the values of
${a \choose b}$ for any $a \in [m]$, $b \in [a]$ in preprocess, so
that we will not need to calculate them again during the process.
\end{itemize}
After doing both changes, $T$ will drop to $\bigT{1}$.

We conclude the discussion with the following Theorem:
\begin{thm} \label{GVB:DET:thmFinal}
Let $q$ be a prime power, $m$ and $k$ positive integers and
$\delta \in [0,1]$. If $k \leq \left(1-H_q(\delta)\right)m$, then
it's possible to construct an $[m,k,\delta m]_q$-LC in time
$\bigT{mq^k}$.
\end{thm}

\section{Conclusion and Open Problems}
We have presented a simple and intuitive construction of linear
codes meeting the GV bound. Our construction is the most efficient
known construction of such linear codes. We used our codes
construction to construct explicitly, in $\bigT{rn \ln{n}}$ time,
very good GT schemes of $\bigT{r^2 \ln{n}}$ tests. It would be
interesting to study whether our linear codes construction can be
made more efficient, or whether it can be improved to construct
better codes. While we managed to close the gap between the sizes
of explicit and non-explicit group testing schemes, the gap in the
important generalization of selectors is still open; closing it is
an interesting and important problem. We believe that other
important special cases of group testing worth studying include
the problem of minimizing the sets accumulative size rather than
their number, and also, solely for algorithmic purposes -- the
case where the tests answers tell not only if there exists an
element in the intersection or not, but rather, how many elements
are there in it.

\addcontentsline{toc}{section}{Bibliography}
\bibliographystyle{plain}
\bibliography{Paper_2}

\end{document}